\documentclass[]{IEEEtran}

\IEEEoverridecommandlockouts

\input epsf
\usepackage{graphicx}
\usepackage{url}
\usepackage{multirow}
\usepackage{xfrac}
\usepackage{bm}

\usepackage{xcolor}   

\DeclareRobustCommand*{\IEEEauthorrefmark}[1]{%
  \raisebox{0pt}[0pt][0pt]{\textsuperscript{\footnotesize\ensuremath{#1}}}}

\begin{document}


 \IEEEpubid{\makebox[\columnwidth]{979-8-3503-6104-9/24/\$31.00~\copyright2024 European Union \hfill} \hspace{\columnsep}\makebox[\columnwidth]{ }}

\title{\LARGE \textcolor{black}{Finite Element Analysis of the Uncertainty Contribution from Mechanical Imperfections in the LNE's Thompson-Lampard Calculable Capacitor}}


\author{\IEEEauthorblockN{Almazbek Imanaliev\IEEEauthorrefmark{1},
Olivier Thevenot\IEEEauthorrefmark{1}, and
Kamel Dougdag\IEEEauthorrefmark{1}} \\
\IEEEauthorblockA{\IEEEauthorrefmark{1}LNE, 29 avenue Roger Hennequin, 78190 Trappes-France\\almazbek.imanaliev@lne.fr}}


\maketitle

\IEEEpubidadjcol

\begin{abstract}

Thompson-Lampard type calculable capacitors (TLCC) serve as electrical capacitance standards, enabling the realization of the farad in the International System of Units (SI) with a combined uncertainty on the order of one part in $10^{8}$. This paper presents an electrostatic finite element (FEM) simulation study focusing on the mechanical imperfections inherent in the developed second generation TLCC at LNE and their influence on the combined uncertainty of the practical realization of the farad. In particular, this study establishes the acceptable tolerances \textcolor{black}{for deviations} from perfect geometrical arrangements of the TLCC electrodes required to achieve the target relative uncertainty of one part in $10^8$.  The simulation predictions are compared with corresponding experimental observations which were conducted with the help of the sub-micron level control of the standard's electrode geometry. \textcolor{black}{In the second generation of the LNE's TLCC, the uncertainty contribution from mechanical imperfections was reduced by at least a factor of 4, as demonstrated by the present FEM analysis. Combined with other improvements, the standard's overall uncertainty meets the target level.}

\end{abstract}


\pagenumbering{gobble}

\section{Introduction}
As outlined in Appendix $2$ of the $9^{th}$ SI Brochure $(2019)$ under the revised SI \cite{bipm_international_2009}, one of the methods of the practical realization of the farad involves the use of a calculable capacitor. Several National Metrology Laboratories, including NIST, NMIA, NRC, NIM\textcolor{black}{, LNE} and BIPM, \textcolor{black}{are involved  in the design and, in many cases, the development of} calculable standards, capable of ensuring the realisation of the farad with a relative uncertainty of $1\textcolor{black}{\times}10^{-7}$ or better, as indicated by the latest international CCEM-$4.2017$ comparison results \cite{gournay2019comparison}.

While the TLCC standards are not dependent conceptually on their physical dimensions \cite{thompson1956new, lampard1957new}, their main limitations stem from the practical implementation challenges. Imperfections in electrode manufacturing and deviations from ideal geometrical arrangements are the main difficulties if uncertainties of less than one part in $10^{7}$ have to be achieved. For a practical Thompson-Lampard capacitor, this implies acceptable tolerances of the order of sub-micrometer on the cylindricity of the intra-electrode geometry and positioning of the movable electrode.

A second generation of the formerly employed calculable capacitor \cite{trapon2003determination}, featuring five cylindrical electrodes initially positioned horizontally at LNE, has been developed with its five main electrodes now oriented vertically \cite{6250948, gournay2011toward, thevenot2010realization}. This setup aims to attain a target uncertainty level of one part in $10^{8}$ \cite{thevenot2023progress, imanaliev2023measuring}. In the new standard, the main improvements are related to the parallelism defects of the electrodes and the coaxiality defect between the capacitor axis and the trajectory of the movable electrode. This paper presents a comprehensive discussion on the finite element analysis and characterization of these two improvements in the new generation Thompson-Lampard capacitor at LNE. 

The numerical analysis derived from this study are broad in scope and can be extended to other standards employing a four-main electrode system.  In the existing literature, a $\text{3-D}$ electric field simulation study \cite{5168646} focused on a calculable capacitor with a four-electrode system, aiming to estimate errors arising from mechanical imperfections such as protruding spikes or step diameter increases in one of the main electrodes. Our study takes a step further by determining uncertainty contributions of the imperfections associated with both the main electrodes and the mobile electrode in a five-electrode system, and subsequently validating them through experimental results.

After briefly introducing the Thomspon-Lampard capacitance theorem extended to five electrodes, the finite element analysis of the mechanical imperfections and their contributions to the uncertainty of the capacitor will be presented. Acceptable tolerances of various geometrical parameters such as the parallelism of the main electrodes and displacement of the movable electrode will be established. Their uncertainty contributions for the second generation TLCC will be reported. Finally, the paper will be concluded with a discussion on perspective.
\IEEEpubidadjcol

\section{TLCC Uncertainty from Imperfections}

\subsection{Perfect TLCC}
\begin{figure}[t]
  \center
  \includegraphics[width=3.0in]{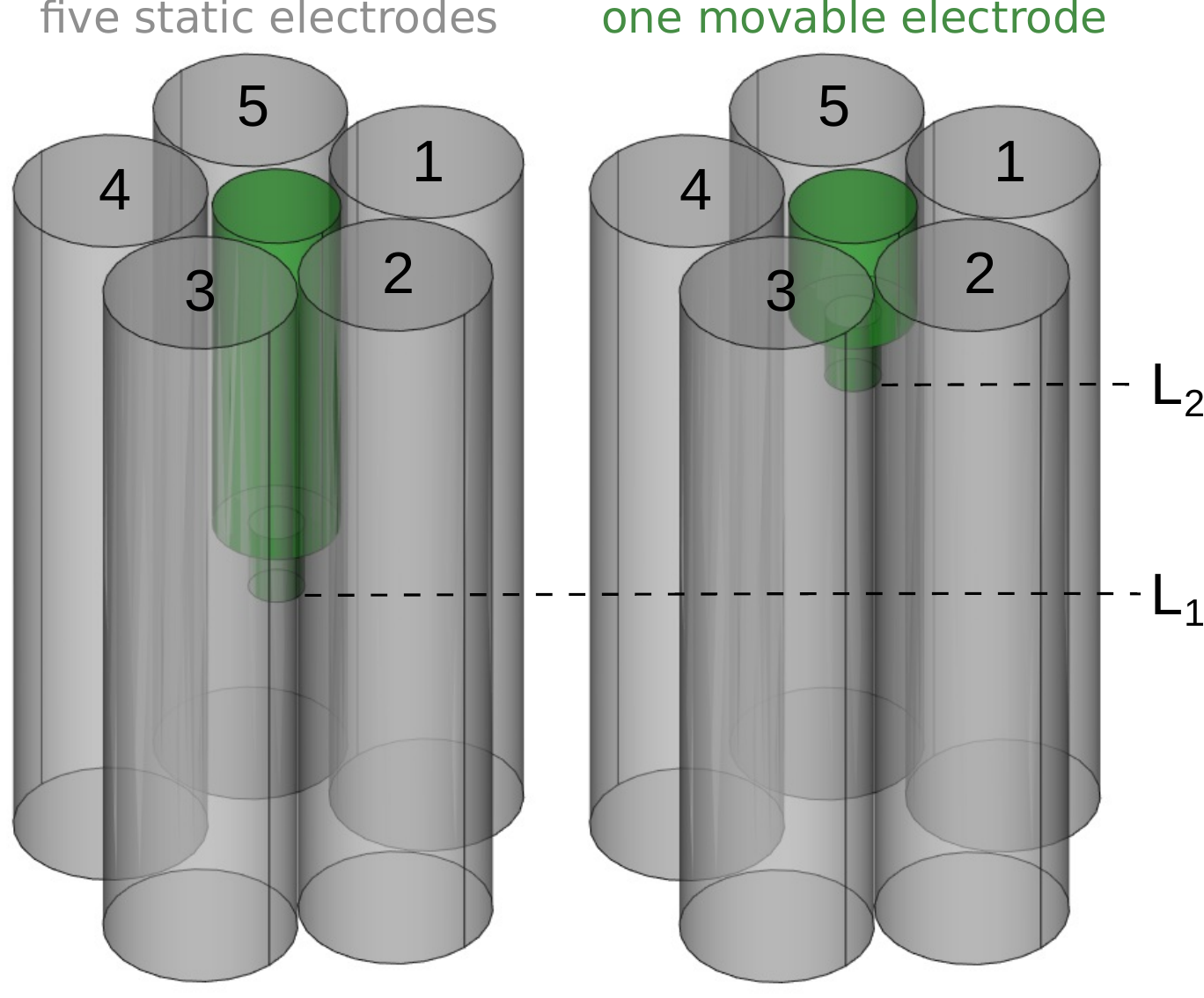}
  \caption{The five static electrodes of the TLCC and its movable guard electrode displacing axially inside the volume enclosed by the five static electrodes.}
  \label{fig:systemFigure}
\end{figure}
As shown in Fig.~\ref{fig:systemFigure}, the TLCC at LNE consists of five static cylindrical electrodes symmetrically positioned at the vertices of a regular pentagon. Additionally, there is a movable guard electrode that displaces axially between the positions $L_{1}$ and $L_{2}$, partially screening the volume enclosed by the static electrodes. To realize the farad in the International System of Units (SI) using the LNE TLCC, a capacitance standard $C$ with a nominal value of $1$~pF is compared to the capacitance variation of the TLCC. This comparison is achieved using a coaxial impedance bridge with a ratio $r=\frac{8}{3}$ \cite{trapon2003determination, thevenot2023progress}. Consequently, the capacitance $C$ is linked to the meter and is expressed as:
\begin{eqnarray}
C & = & 2r\overline{\gamma}\Delta L. \label{eq1pF}
\end{eqnarray}
Here, $\Delta L = L_{1} - L_{2}$, and the mean capacitance per unit length $\overline{\gamma}$ is defined as:
\begin{eqnarray}
\overline{\gamma} & = & \frac{\gamma_{2,5}+\gamma_{1,3}+\gamma_{2,4}+\gamma_{3,5}+\gamma_{1,4}}{5} \label{eqgaverage}
\end{eqnarray}
 where $\gamma_{i,j}$ represents the capacitance per unit length between the static electrodes $i$ and $j$. The factor of $2$ arises from the measurement protocol, where it is chosen to electrically link two static electrodes. Experimentally, $r$ and $\Delta L$ are directly accessible with sufficient accuracy, unlike $\overline{\gamma}$. To maintain the validity of equation (\ref{eq1pF}), it is essential that the edge electric fields remain undisturbed during the displacement of the movable guard between the two positions $L_{1}$ and $L_{2}$.

For an ideal TLCC, the Lampard's theorem \cite{lampard1957new} establishes the relationships between individual capacitances per unit length, $\gamma_{i,j}$, as follows:
\begin{eqnarray}
e^{-\frac{\pi}{\epsilon_{0}}(\gamma_{1,3}+\gamma_{1,4})}+ e^{-\frac{\pi}{\epsilon_{0}}\gamma_{2,5}} & = &  1 \nonumber \\
e^{-\frac{\pi}{\epsilon_{0}}(\gamma_{2,4}+\gamma_{2,5})}+ e^{-\frac{\pi}{\epsilon_{0}}\gamma_{1,3}} & = &  1 \nonumber \\
e^{-\frac{\pi}{\epsilon_{0}}(\gamma_{3,5}+\gamma_{1,3})}+ e^{-\frac{\pi}{\epsilon_{0}}\gamma_{2,4}} & = &  1 \label{eqlampard} \\
e^{-\frac{\pi}{\epsilon_{0}}(\gamma_{1,4}+\gamma_{2,4})}+ e^{-\frac{\pi}{\epsilon_{0}}\gamma_{3,5}} & = &  1 \nonumber \\
e^{-\frac{\pi}{\epsilon_{0}}(\gamma_{2,5}+\gamma_{3,5})}+ e^{-\frac{\pi}{\epsilon_{0}}\gamma_{1,4}} & = &  1 \nonumber
\end{eqnarray}
where $\epsilon_{0}$ represents the vacuum permittivity. Moreover, in the case of a perfectly symmetrical TLCC geometry, the capacitances per unit length, as determined by these relations, are given by:
\begin{eqnarray}
\overline{\gamma} & = & \gamma_{i,j} \nonumber \\ 
                  & = & \gamma_{0} = \frac{\epsilon_{0}}{\pi} \ln \frac{2}{\sqrt{5}-1} = 1.35623... \text{ pF/m.} \label{eqgsymm}
\end{eqnarray}
The capacitance value $C$ is subsequently derived from equations (\ref{eq1pF}) and (\ref{eqgsymm}), leading to the expression $C = 2r\gamma_{0}\Delta L$. In practice, the effort will be made to approach as close as possible to this ideal situation when developing a new TLCC.

\subsection{Imperfect TLCC}
In real-world scenarios, the TLCC is subject to imperfections, which include slight deviations in the physical dimensions of the electrode bars, as well as variations in their alignments and positions. These imperfections introduce deviations in the value of $\overline{\gamma}$ from its theoretical value $\gamma_{0}$, expressed as:
\begin{eqnarray}
\overline{\gamma} & = & \gamma_{0} + \Delta \overline{\gamma}.
\end{eqnarray}
Consequently, these deviations impact the capacitance $C$ through equation (\ref{eq1pF}). Since the precise value of $\Delta\overline{\gamma}$ cannot be experimentally determined with sufficient accuracy, it is instead considered as part of the uncertainty in $\overline{\gamma}$ and, consequently, in $C$. The relative combined variance of $C$ \cite{JCGMGUM} is given by
\begin{eqnarray}
\left[ \frac{u(C)}{C} \right]^{2} & = & \left[ \frac{u(r)}{r} \right]^{2}+\left[ \frac{u(\overline{\gamma})}{\gamma_{0}} \right]^{2}+\left[ \frac{u(\Delta L)}{\Delta L} \right]^{2}
\end{eqnarray}
The relative standard type B uncertainties of the three terms were estimated as follows: $\frac{u(r)}{r}=1.64\textcolor{black}{\times}10^{\textcolor{black}{-8}}$, $\frac{u(\overline{\gamma})}{\gamma_{0}} =3.85\textcolor{black}{\times}10^{\textcolor{black}{-8}}$ and $\frac{u(\Delta L)}{\Delta L} =0.09\textcolor{black}{\times}10^{\textcolor{black}{-8}}$ in the first generation TLCC \cite{trapon2003determination}. In this paper, our focus lies in determining the dominant relative uncertainty, $\frac{u(\overline{\gamma})}{\gamma_{0}}$, which quantifies the contribution of the calculable standard imperfections to the overall combined relative standard uncertainty of $C$. \textcolor{black}{For information on other uncertainty contributions and their improvements in the second generation TLCC, refer to \cite{thevenot2023progress}}.

When considering imperfections in the TLCC, it's crucial to distinguish between two types. The first type of imperfections alters the individual capacitance per unit length $\gamma_{i,j}$ but does not impact their mean value $\overline{\gamma}$. Additionally, the relations given in (\ref{eqlampard}) remain applicable for this type. Therefore, these imperfections, such as non-symmetrical positioning of the static electrodes or differences in their radii, do not introduce uncertainty to the value of $C$.

On the contrary, the second type of imperfections\textcolor{black}{, such as parallelism errors or trajectory errors (described below),} which impact the value of $\overline{\gamma}$, become crucial during the TLCC assembly and contribute to the uncertainty in determining $C$. These imperfections introduce residual errors in the Lampard's relations provided in (\ref{eqlampard}):
\begin{eqnarray}
e_{1} & = & e^{-\frac{\pi}{\epsilon_{0}}(\gamma_{1,3}+\gamma_{1,4})}+ e^{-\frac{\pi}{\epsilon_{0}}\gamma_{2,5}} - 1 \nonumber \\
e_{2} & = & e^{-\frac{\pi}{\epsilon_{0}}(\gamma_{2,4}+\gamma_{2,5})}+ e^{-\frac{\pi}{\epsilon_{0}}\gamma_{1,3}} -  1 \nonumber \\
e_{3} & = & e^{-\frac{\pi}{\epsilon_{0}}(\gamma_{3,5}+\gamma_{1,3})}+ e^{-\frac{\pi}{\epsilon_{0}}\gamma_{2,4}} -  1 \label{eqlamparderr} \\
e_{4} & = & e^{-\frac{\pi}{\epsilon_{0}}(\gamma_{1,4}+\gamma_{2,4})}+ e^{-\frac{\pi}{\epsilon_{0}}\gamma_{3,5}} -  1 \nonumber \\
e_{5} & = & e^{-\frac{\pi}{\epsilon_{0}}(\gamma_{2,5}+\gamma_{3,5})}+ e^{-\frac{\pi}{\epsilon_{0}}\gamma_{1,4}} -  1 \nonumber
\end{eqnarray}
\textcolor{black}{As demonstrated in the appendix,} a relationship between the mean error $\overline{e} = \sum_{i} e_{i}/5$ of these five residual errors $e_{i}$ and $\frac{u(\overline{\gamma})}{\gamma_{0}}$ can be found:
\begin{eqnarray}
\overline{e}    & = & -\ln \frac{2}{\sqrt{5}-1} (1+e^{-2\ln \frac{2}{\sqrt{5}-1} })\frac{u(\overline{\gamma})}{\gamma_{0}} \nonumber \\
                & = & -\frac{1}{\kappa}\frac{u(\overline{\gamma})}{\gamma_{0}} \label{eqganderelation}
\end{eqnarray}
where $\kappa \approx 1.5$.
\begin{figure}[t]
  \center
  \includegraphics[width=3.0in]{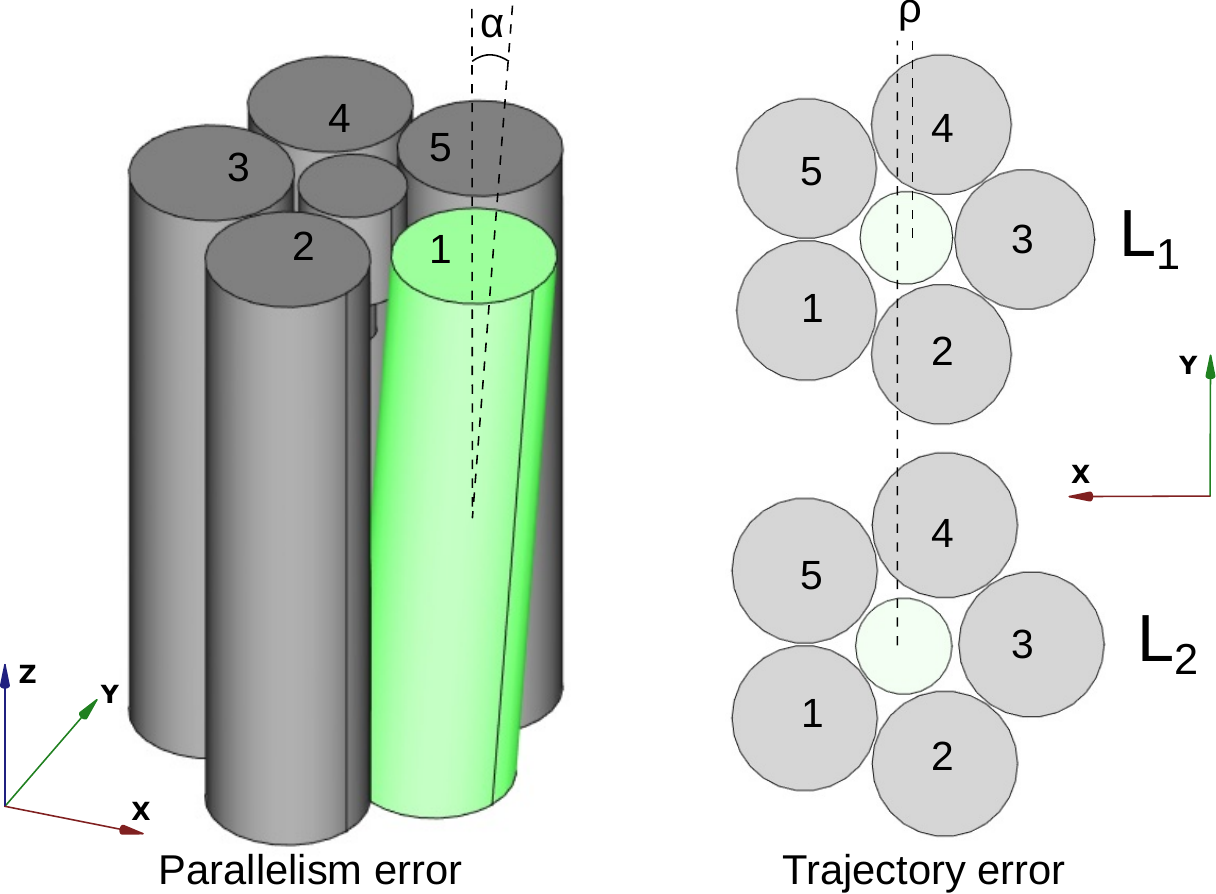}
  \caption{TLCC imperfections: parallelism error (left) due to the tilt $\alpha$ of the first electrode and the trajectory error due to the lateral displacement $\rho$ between axial positions $L_{1}$ and $L_{2}$ of the movable electrode.}
  \label{fig:imperfectFigure}
\end{figure}

In the following discussion, attention is directed towards addressing the two predominant imperfections highlighted in the previous TLCC at LNE, as studied by Trapon et al. \cite{trapon2003determination}. These imperfections are visually represented in Fig.~\ref{fig:imperfectFigure} and relate to the parallelism of the static electrodes and the trajectory error of the movable electrode. To characterize these imperfections, let the parallelism error be quantified in terms of the tilt angle $\alpha$ of one of the static electrodes, and the trajectory error be expressed in terms of the lateral displacement $\rho$ of the movable electrode. The relative uncertainty $\frac{u(\overline{\gamma})}{\gamma_{0}}$ can be evaluated from:
\begin{eqnarray}
\left[ \frac{u(\overline{\gamma})}{\gamma_{0}} \right]^{2} & = & \left[ \frac{1}{\gamma_{0}}\frac{\partial \overline{\gamma}}{\partial \alpha}u(\alpha) \right]^{2}+\left[ \frac{1}{\gamma_{0}}\frac{\partial \overline{\gamma}}{\partial \rho}u(\rho) \right]^{2} \label{eqsenscoeff1}\\
& = & \left[-\kappa \frac{\partial \overline{e}}{\partial \alpha}u(\alpha) \right]^{2}+\left[ -\kappa\frac{\partial \overline{e}}{\partial \rho}u(\rho) \right]^{2} \label{eqsenscoeff2}
\end{eqnarray}
Essentially, the sensitivity coefficients $\frac{\partial \overline{\gamma}}{\partial \alpha}$, $\frac{\partial \overline{\gamma}}{\partial \rho}$, or $\frac{\partial \overline{e}}{\partial \alpha}$, $\frac{\partial \overline{e}}{\partial \rho}$ specify how the output estimates of $\overline{\gamma}$ or $\overline{e}$ vary with changes in the values of the input estimates $\alpha$ and $\rho$. If these sensitivity coefficients are known, the relative uncertainty $\frac{u(\overline{\gamma})}{\overline{\gamma}}$ can be readily determined. However, it is noteworthy that these sensitivity coefficients may not always be analytically known or experimentally accessed. In such cases, the finite element simulation approach emerges as a valuable tool to determine these coefficients effectively and thereby assess the relative uncertainty in the TLCC standard.  It will be elaborated in the following section.

\section{FEM Simulation}
\begin{figure}[t]
  \center
  \includegraphics[width=3.0in]{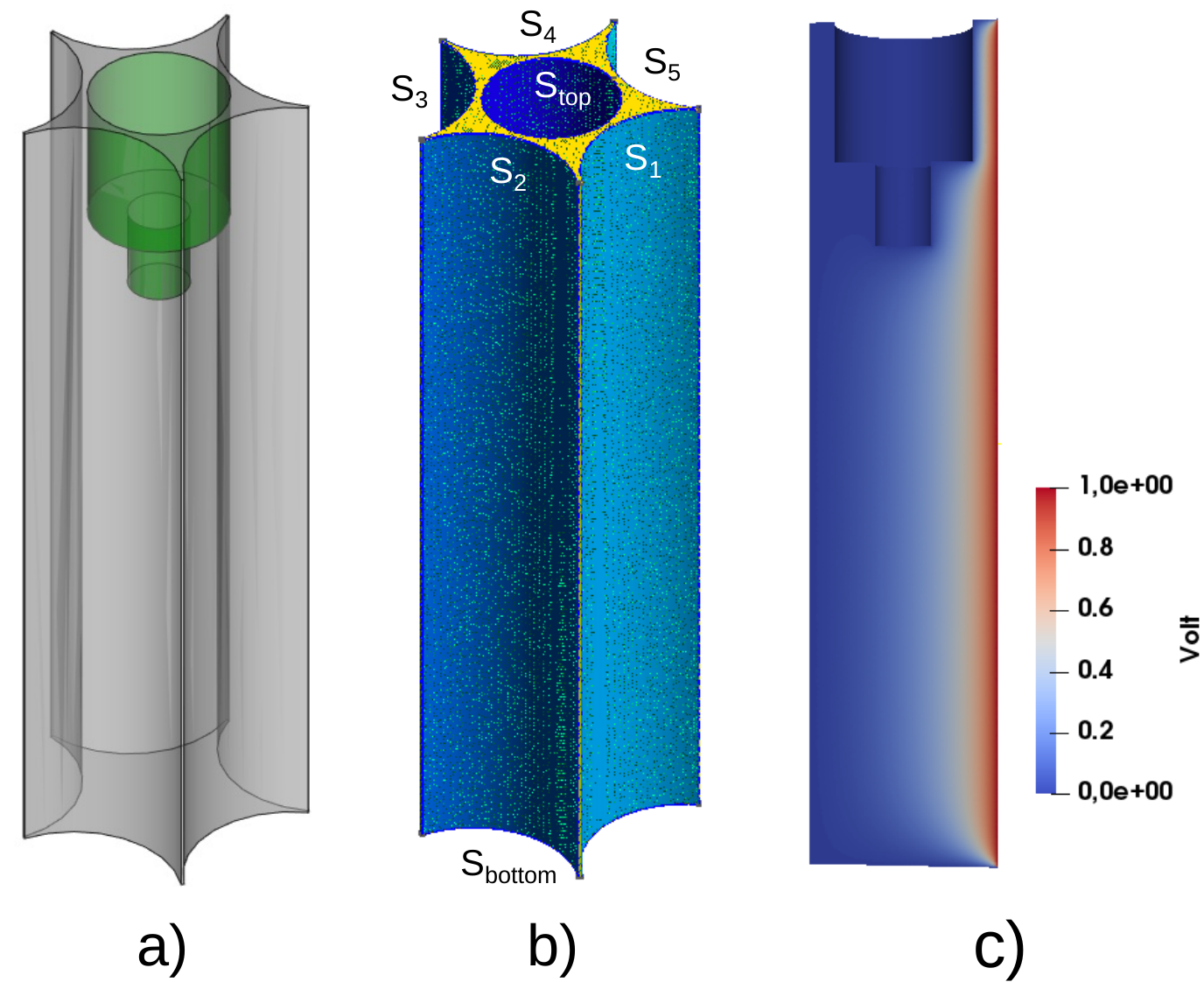}
  \caption{a) volume used for the FEM simulation b) meshing of the volume with surfaces $S_{i}$ c) evaluated potential function at the axial cut of the volume with the boundary conditions $\phi_{S_{1}}=1$~V and $0$~V elsewhere.}
  \label{fig:cavityFigure}
\end{figure}

The FEM simulation involves the resolution of a three-dimensional (3D) electrostatic problem within the volume delimited by the electrodes, depicted in panel a) of Fig.~\ref{fig:cavityFigure}. In the initial phase of the simulation, the volume is created using computer-aided design (CAD) software, specifically FreeCAD~\cite{noauthor_freecad_nodate}. The dimensions of this volume are aligned with the actual standard dimensions of the TLCC, ensuring that the simulation input closely mirrors real-world conditions. During the simulations, the length of the volume in Fig.~\ref{fig:cavityFigure} is varied from $200$~mm to $300$ mm. \textcolor{black}{This range represents a trade off between minimizing the fringing effects, which sets the lower limit, and managing the number of mesh elements, which sets the upper limit.} The radius of the circumscribing circle for the volume's cross-section is approximately $65$ mm. 

Subsequently, the chosen volume undergoes meshing using a three-dimensional finite element mesh generator, Gmsh \cite{geuzaine2009gmsh}, as depicted in b) of Fig.~\ref{fig:cavityFigure}. To improve the accuracy of simulation results, the mesh size needs to be reduced. This results in a high total mesh number for the volume with real-world dimensions. However, a high total mesh number not only demands substantial hardware memory but also extends the simulation duration. The mesh size of $0.5$~mm was selected to perform computations on a personal computer equipped with a $16$~GB memory and an Intel i7 processor, ensuring that the calculations are completed within a reasonable time frame. To ensure the sensitivity coefficients in (\ref{eqsenscoeff1}) and (\ref{eqsenscoeff2}) are sufficiently accurate in simulations, the imperfections of the TLCC were intentionally magnified beyond their actual magnitude, amplifying their impact for more pronounced effects.

The next stage involves the resolution of the Poisson equation,
\begin{eqnarray}
\nabla^{2}\phi & = & 0  \label{eqpoisson}
\end{eqnarray}
to determine the electrostatic potential. The boundary conditions $\phi_{S_{i}}$ are specified on the surfaces of the defined volume. This numerical solution is accomplished using the finite element method, implemented through the open-source computing platform FEniCS \cite{AlnaesBlechta2015a}. An illustration of the resulting potential function, which is obtained by solving Poisson's equation (\ref{eqpoisson}) with the Dirichlet boundary conditions: $\phi_{S_{1}}=1$ V, $\phi_{S_{2,3,4,5, \text{top}, \text{bottom}}}=0$ V, is displayed in c) of Fig.~\ref{fig:cavityFigure}. Then, the capacitance $C_{1,3}$ is derived from
\begin{eqnarray}
C_{1,3} & = & \frac{\int_{S_{3}}-\nabla\phi\cdot\mathbf{n} \,dS}{\phi_{S_{1}}-\phi_{S_{3}}}\epsilon_{0}
\end{eqnarray}
This expression represents the ratio of surface charge to the potential difference between the corresponding surfaces. Similarly, the remaining four capacitances: $C_{2,4}$, $C_{3,5}$, $C_{1,4}$ and $C_{2,5}$ are calculated by systematically varying the boundary conditions for each iteration of the simulation. 

The capacitances $C_{i,j}$ have been computed for a specific position $L_{i}$ of the movable electrode in the previous step. Finally, by conducting preceding calculations iteratively with different values of $L_{i}$, it becomes possible to compute the capacitances per unit length $\gamma_{i,j}$ and their mean value $\overline{\gamma}$ in equation (\ref{eqgaverage}). The sensitivity coefficients can then be determined by assessing the variations in $\overline{\gamma}$ or $\overline{e}$ in response to changes in the parameters $\alpha$ and $\rho$, which were introduced previously. These parameters are applied at the initial stage while creating the volume with the CAD software. This approach essentially forms the basis for the simulations presented in this paper. 
\begin{table}[h]
\caption{Simulation of the ideal TLCC standard \label{tab:table0}}
\renewcommand\arraystretch{1.3}
\centering
\begin{tabular}{|c||c|c|}
\hline
\multirow{2}{*}{Relative errors} & \multirow{2}{*}{Values \textcolor{black}{$\times 10^{6}$}}  & Corrected values \\
& & = Values \textcolor{black}{$\times 10^{6}$}- $193.1$\\
\hline\hline
$\sfrac{\gamma_{2,5}-\gamma_{0}}{\gamma_{0}}$  & -225.9 & -32.8\\

$\sfrac{\gamma_{1,3}-\gamma_{0}}{\gamma_{0}}$   & -175.9 & 17.2\\

$\sfrac{\gamma_{2,4}-\gamma_{0}}{\gamma_{0}}$  & -197.7 & -4.6\\

$\sfrac{\gamma_{3,5}-\gamma_{0}}{\gamma_{0}}$   & -196.1 & -3.0\\

$\sfrac{\gamma_{1,4}-\gamma_{0}}{\gamma_{0}}$   & -169.9 & 23.2\\
\hline\hline
$\sfrac{\overline{\gamma}-\gamma_{0}}{\gamma_{0}}$ & -193.1 & 0.0\\

$-\kappa\overline{e}$ & -193.7 & -0.6\\
\hline
\end{tabular}
\end{table}
A simulation of the ideal and symmetrical TLCC standard yields relative errors in the capacitances per unit length, as summarized in Table~\ref{tab:table0}. Ideally, all values in this table should be zero, in accordance with equation (\ref{eqgsymm}). However, an offset error of $-193.1$\textcolor{black}{$\times 10^{-6}$} is observed, closely associated with the spacings of $1$~mm between the static electrodes and the mesh size of $0.5$~mm. By reducing both of these parameters in simulations, the offset error was diminished. It is worth to highlight that Neumann boundary conditions, where the normal derivative of the potential function is set to zero, are enforced on these spacings during the simulation. Since this offset error remains constant and is known, it does not affect the accuracy of the sensitivity coefficients to be computed in subsequent sections. The relevant uncertainty contribution from the FEM simulation is therefore represented by the values reported in the last column of Table~\ref{tab:table0}, approximately $20$\textcolor{black}{$\times 10^{-6}$}.

\section{Parallelism error}
In this section, we will determine the sensitivity coefficients $\frac{\partial \overline{\gamma}}{\partial \alpha}$ and $\frac{\partial \overline{e}}{\partial \alpha}$. To achieve this, we tilt the static electrode no. $1$ by an angle $\alpha$, as illustrated in Fig.~\ref{fig:imperfectFigure}. The variations in $\overline{\gamma}$ and $\overline{e}$ are then computed through FEM simulations, as detailed in the preceding section.

An influential factor affecting these sensitivity coefficients is the presence of the spike, as discussed in \cite{trapon2003determination}. The spike is represented by a cylinder with a smaller radius located at the bottom part of the movable guard, as shown in Fig.~\ref{fig:systemFigure} and Fig.~\ref{fig:cavityFigure}. We will analyze and compare the results with and without the presence of the spike separately in the subsequent discussions.

\subsection{Electrode tilt without spike} \label{ssec:num1}
\begin{figure}[t]
  \center
  \includegraphics[width=3.4in]{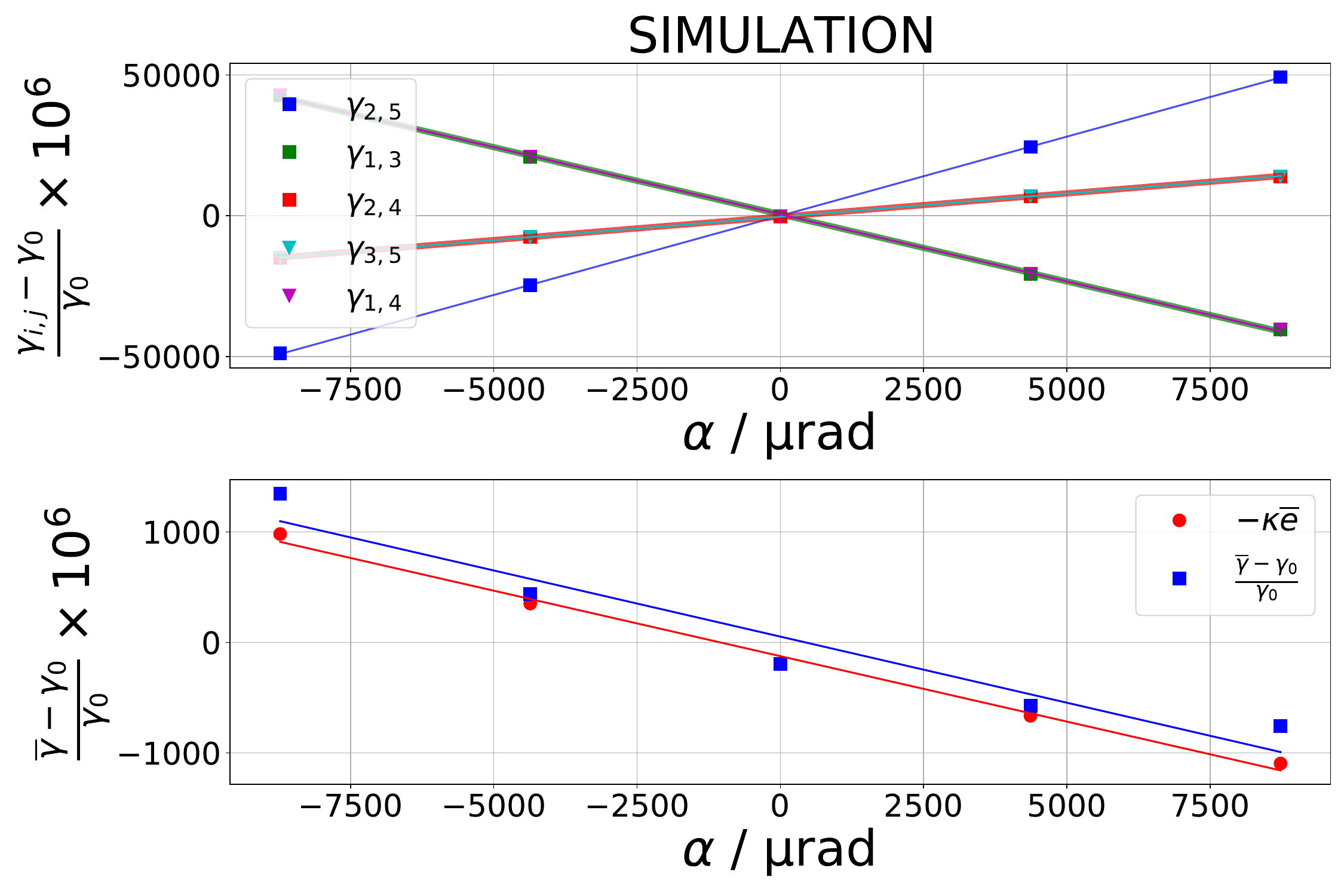}
  \caption{Simulation results, depicting the relative variations of individual capacitances per unit length (shown in the top plot) and of their average value (depicted in the bottom plot) with respect to the tilt angle $\alpha$ of the first static electrode.}
  \label{fig:tiltnospikeFigure}
\end{figure}

The simulation results of $\frac{\gamma_{i,j}-\gamma_{0}}{\gamma_{0}}$ and $\frac{\overline{\gamma}-\gamma_{0}}{\gamma_{0}}$ are plotted in Fig.~\ref{fig:tiltnospikeFigure} as a function of the tilt angle $\alpha$.
According to these plots, the variations in $\gamma_{i,j}$ are more pronounced than their average value. The relative difference $\frac{\overline{\gamma} - \gamma_{0}}{\gamma_{0}}$ were computed using both equation (\ref{eqgaverage}) and (\ref{eqganderelation}).
\begin{figure}[t]
  \center
  \includegraphics[width=3.4in]{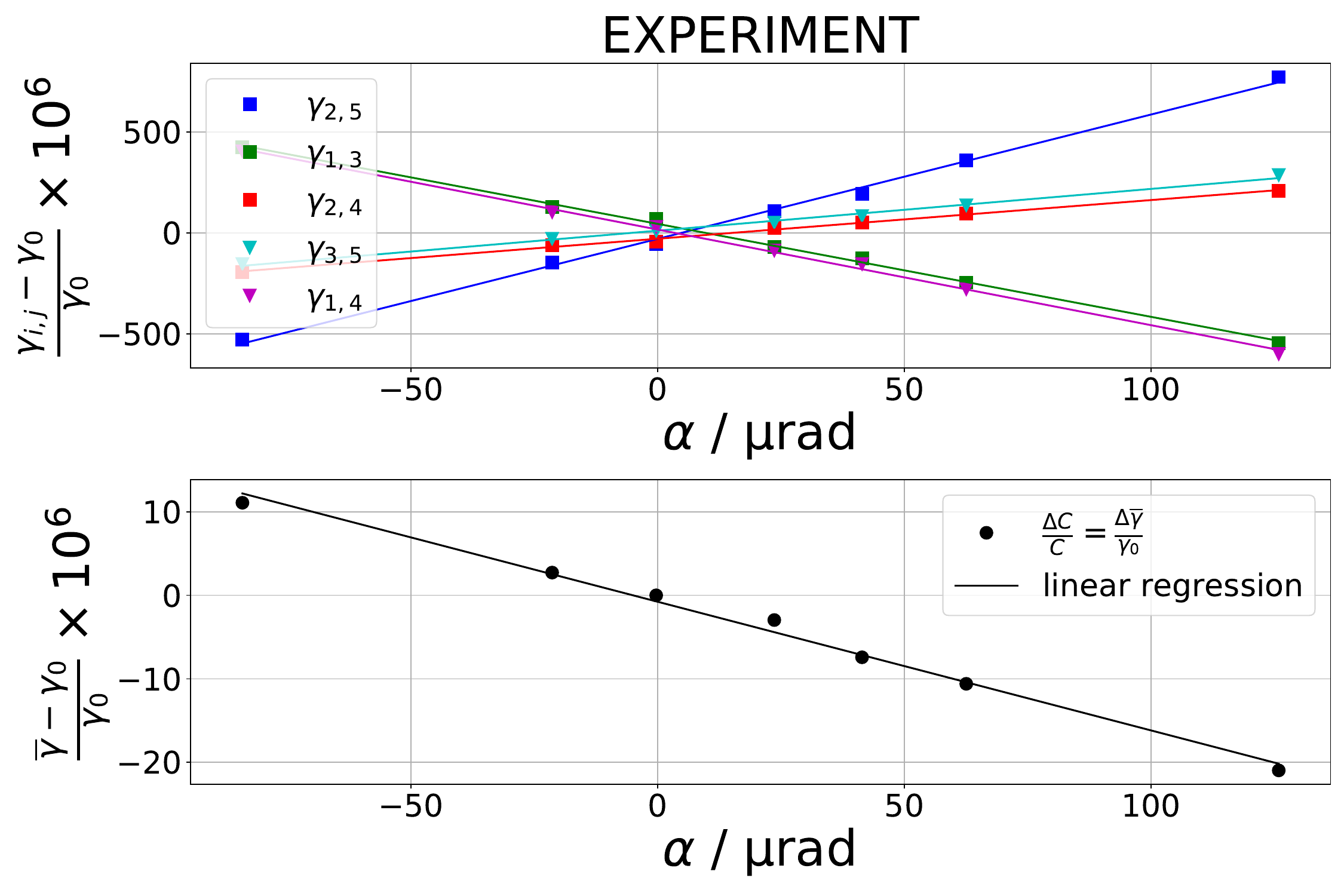}
  \caption{\textcolor{black}{Experimental results, showing the relative variations of individual capacitances per unit length (shown in the top plot) and of their average value (shown in the bottom plot) with respect to the tilt angle $\alpha$ of the first static electrode.
  They are reproduced from~\cite{trapon2003determination} and to be compared with the simulation results in Fig.~\ref{fig:tiltnospikeFigure}.}}
  \label{fig:exptiltnospikeFigure}
\end{figure}
To validate the obtained simulation results, we compare them with corresponding experimental findings from studies conducted by Trapon et al.~\cite{trapon2003determination}. The reproduced experimental results are presented in Fig.~\ref{fig:exptiltnospikeFigure}. A comparison of the determined sensitivity coefficients from both the simulation and the experiment is given in Table~\ref{tab:table1}. They were derived through linear regression analysis performed on the data presented in Fig.~\ref{fig:tiltnospikeFigure} and Fig.~\ref{fig:exptiltnospikeFigure}. The uncertainties enclosed within parentheses come exclusively from regression uncertainties. The signal-to-noise ratio of the sensitivity coefficients calculated using equation (\ref{eqganderelation}) is approximately four times higher than that achieved with equation (\ref{eqgaverage}) in simulations. For subsequent analysis, this method of calculation is adopted to determine the sensitivity coefficients. The reported values in Table~\ref{tab:table1} demonstrate good agreement, and small discrepancies can be explained by examining the experimental results.
\begin{table}[h]
\caption{Sensitivity coefficients of the parallelism error (no spike) \label{tab:table1}}
\renewcommand\arraystretch{1.3}
\centering
\begin{tabular}{|c||c|c|}
\hline
\multirow{2}{*}{Sensitivity coefficients} & Simulated  & Experimental \\
 & \textcolor{black}{$\times 10^{6}$~µrad}  & \textcolor{black}{$\times 10^{6}$~µrad} \\
\hline\hline
$\sfrac{\partial \gamma_{2,5}}{\gamma_{0}\partial \alpha}$ & 5.62(2) & 6.16(15)\\

$\sfrac{\partial \gamma_{1,3}}{\gamma_{0}\partial \alpha}$ & -4.77(5) & -4.61(10)\\

$\sfrac{\partial \gamma_{2,4}}{\gamma_{0}\partial \alpha}$ & 1.66(1) & 1.92(5)\\

$\sfrac{\partial \gamma_{3,5}}{\gamma_{0}\partial \alpha}$ & 1.66(1) & 2.07(7)\\

$\sfrac{\partial \gamma_{1,4}}{\gamma_{0}\partial \alpha}$ & -4.77(5) & -4.74(11)\\
\hline\hline
$\sfrac{\partial \overline{\gamma}}{\gamma_{0}\partial \alpha}$ & -0.120(19) & \multirow{2}{*}{-0.154(6)}\\
$-\kappa\sfrac{\partial \overline{e}}{\partial \alpha}$ &  -0.118(5) & \\
\hline
\end{tabular}
\end{table}
Notably, asymmetry is observed among the sensitivity coefficients $\frac{\partial \gamma_{i,j}}{\partial \alpha}$ in the experimental data. This suggests the possibility that the orientation of electrode no.~$1$ during tilting deviated slightly from the intended radial outward position. Additionally, errors in the determined sensitivity coefficients from both simulations and experiments contribute to the disparities observed in the comparison.

\subsection{Electrode tilt with spike}
\begin{figure}[b]
  \center
  \includegraphics[width=3.4in]{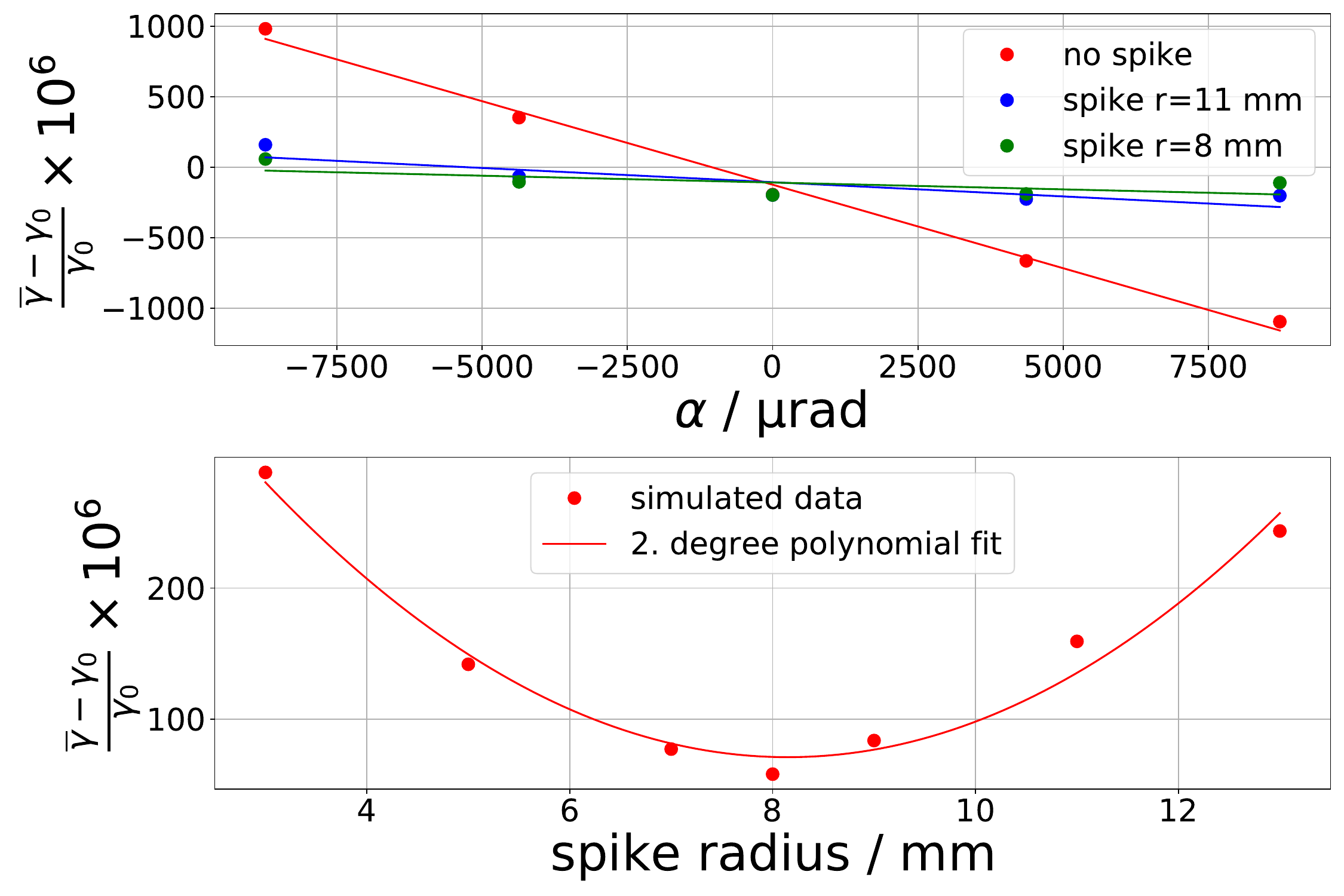}
  \caption{(Top plot) Simulated relative changes in $\overline{\gamma}$ with varying tilt angles $\alpha$ for three different spike radii (Bottom plot) Relative changes in $\overline{\gamma}$ with fixed tilt angle $\alpha=-8726$~µrad and varying spike radii}
  \label{fig:spikeoptimFigure}
\end{figure}
The spike redirects current between two static electrodes to the ground, thus altering the mean capacitance per unit length $\overline{\gamma}$. This change is in addition to the variations induced by the parallelism defect, as simulated in the previous subsection~\ref{ssec:num1}. The combined effects of these alterations can oppose each other, potentially counterbalancing the overall $\overline{\gamma}$ value, particularly under specific spike dimensions. 

The simulation results in Fig.~\ref{fig:spikeoptimFigure} reveal that varying spike radii enables the minimization of relative changes in $\overline{\gamma}$, effectively reducing uncertainty contributions from parallelism errors. The height of the spike was fixed to $30$~mm for these simulations. The optimal spike configuration, with a radius of $8$~mm, is identified from the bottom plot of Fig.~\ref{fig:spikeoptimFigure}. The sensitivity coefficient, determined with this optimal spike radius in the upper plot, is $\frac{1}{\gamma_{0}}\frac{\partial \overline{\gamma}}{\partial \alpha} = -0.0097$\textcolor{black}{$\times 10^{-6}$~µrad$^{-1}$}, resulting in a significant improvement factor of $12$ compared to the system without a spike. In the study by Trapon et al.~\cite{trapon2003determination}, authors reported an improvement factor of $25$ which we find to be inaccurate \textcolor{black}{due to a calculation error involving an erroneous multiplication by a constant $\kappa$ in equation (\ref{eqganderelation})}. The corrected value falls within the range of $14$ to $17$, aligning more closely with the simulation results.

Employing the determined sensitivity coefficient $\frac{1}{\gamma_{0}}\frac{\partial \overline{\gamma}}{\partial \alpha} = -0.0097 $\textcolor{black}{$\times 10^{-6}$~µrad$^{-1}$} and equation (\ref{eqsenscoeff1}), a constraint on the tilt angle $\alpha$ can be defined to maintain the uncertainty contribution from parallelism error below one part in $10^8$. By considering the uncertainties in the tilt angles of the five static electrodes as uncorrelated, the accepted tolerance for the tilt angle on every electrode is given by:
\begin{eqnarray}
u(\alpha) & < &  \frac{10^{-8}}{\sqrt{5}|\frac{1}{\gamma_{0}}\frac{\partial \overline{\gamma}}{\partial \alpha} |} \approx 0.46~\text{µrad} \label{eq:alphatol}
\end{eqnarray}

\section{Trajectory error}

\subsection{Displacement of the movable electrode} \label{ssec:num2}
\begin{figure}[t]
  \center
  \includegraphics[width=3.4in]{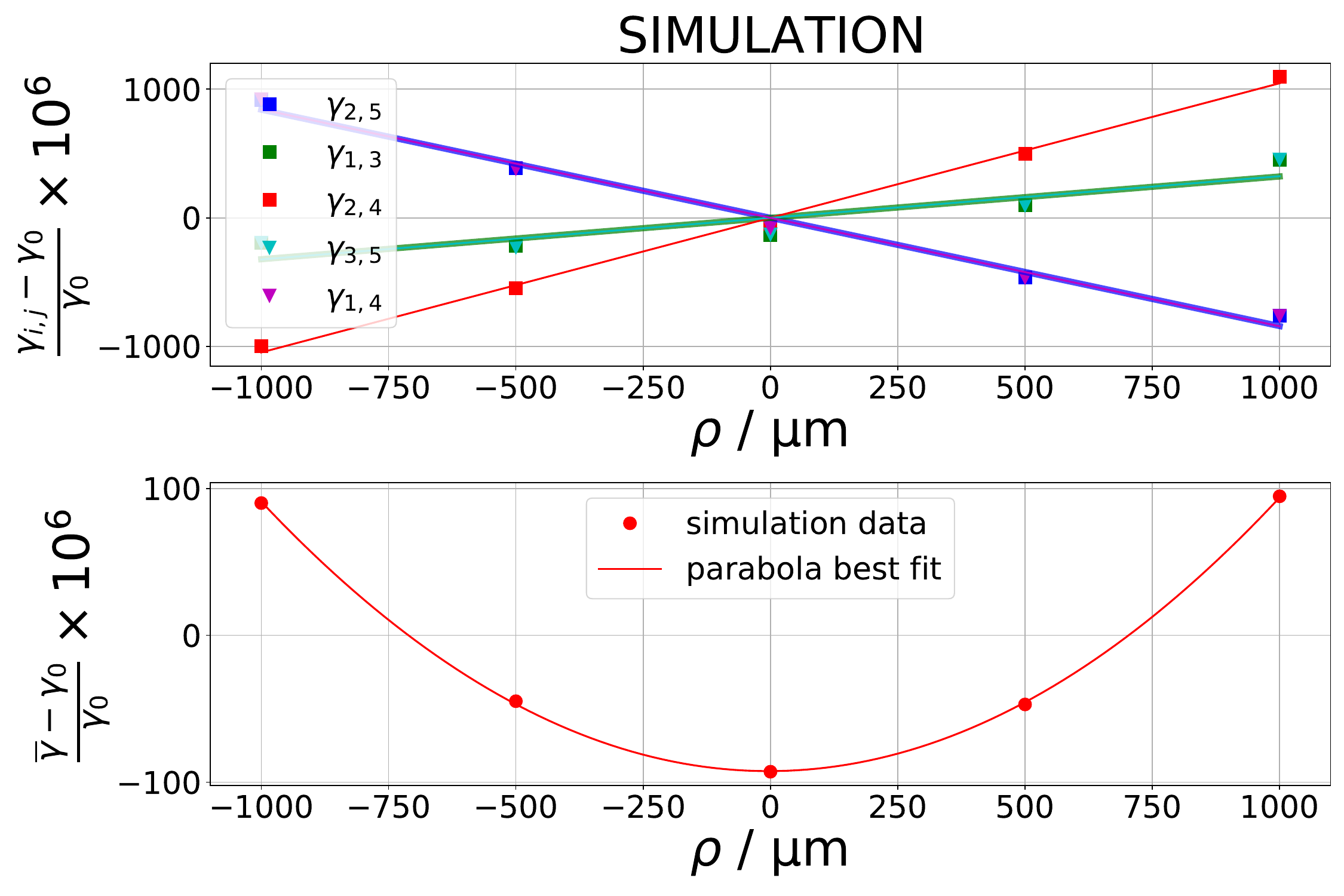}
  \caption{(Top plot) Simulated relative changes in $\gamma_{i,j}$ with varying movable electrode displacement $\rho$ along x-axis, and (Bottom plot) relative changes in their average value.}
  \label{fig:trajerrsimFigure}
\end{figure}
\begin{figure}[t]
  \center
  \includegraphics[width=3.4in]{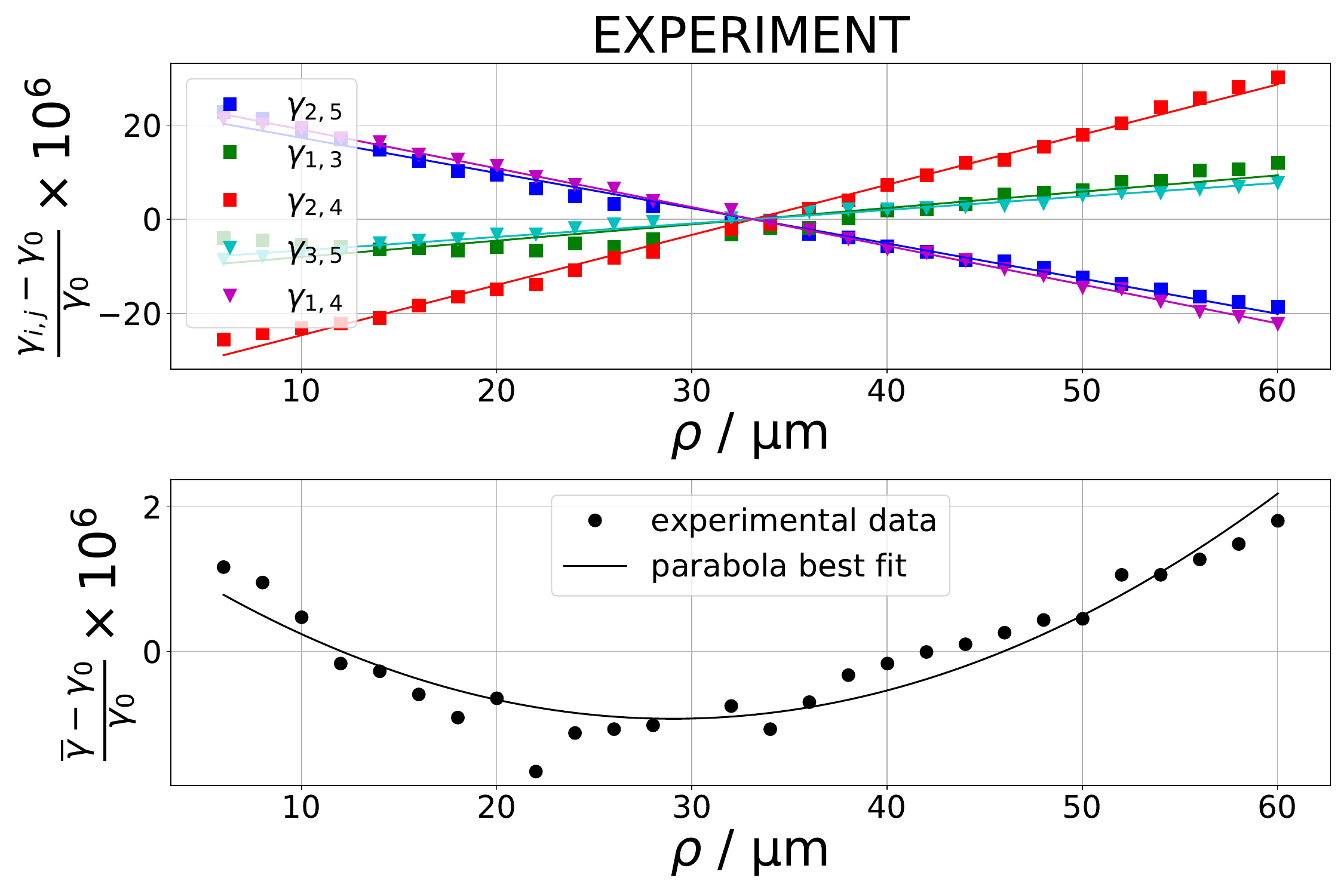}
  \caption{\textcolor{black}{(Top plot) Experimental relative changes in $\gamma_{i,j}$ with varying movable electrode displacement $\rho$ along x-axis, and (Bottom plot) relative changes in their average value. They are obtained with the setup in~\cite{thevenot2023progress}. A discrepancy of a factor of 18 was found in relative variation of $\overline{\gamma}$ when compared to the simulation results in Fig.~\ref{fig:trajerrsimFigure}. For explanation refer to the subsection~\ref{ssec:num3}.}}
  \label{fig:trajerrexpFigure}
\end{figure}
In this section, we will examine how the uncertainty is affected by errors in the trajectory of the movable electrode. Specifically, we displaced the movable electrode horizontally by a distance $\rho$ from the static electrode $3$ (along the x-axis) between positions $L_{1}$ and $L_{2}$ as shown in  Fig.~\ref{fig:imperfectFigure}. The results from simulations and experiments on the relative variation in $\gamma_{i,j}$ and $\overline{\gamma}$ can be seen in Fig.~\ref{fig:trajerrsimFigure} and Fig.~\ref{fig:trajerrexpFigure}. The experimental data isn't centered at $\rho=0$ because it reflects the displacement readings of the piezo-actuators, which might not align precisely with the position of the movable electrode at the midpoint of the five static electrodes. For more information on the sub-micron level control and measurement of the lateral position of the movable electrode in the new TLCC standard at LNE, refer to \cite{thevenot2023progress}. In contrast to the linear relationship observed in parallelism error, the behavior of $\overline{\gamma}$ exhibits a quadratic pattern with respect to $\rho$ in the case of trajectory error. The corresponding sensitivity coefficients were computed thus using a parabolic regression on the data in Fig.~\ref{fig:trajerrsimFigure} and Fig.~\ref{fig:trajerrexpFigure}. The determined sensitivity coefficients for both experimental and simulation data are presented in Table~\ref{tab:table2}.
\begin{table}[h]
\caption{Sensitivity coefficients of the trajectory error ($\rho$) \label{tab:table2}}
\renewcommand\arraystretch{1.3}
\centering
\begin{tabular}{|c||c|c|}
\hline
\multirow{2}{*}{Sensitivity coefficients} & Simulated  & Experimental \\
 & \textcolor{black}{$\times 10^{6}$~µm}  & \textcolor{black}{$\times 10^{6}$~µm} \\
\hline\hline
$\sfrac{\partial \gamma_{2,5}}{\gamma_{0}\partial \rho}$ & -0.84 & -0.75\\

$\sfrac{\partial \gamma_{1,3}}{\gamma_{0}\partial \rho}$ & 0.32 & 0.34\\

$\sfrac{\partial \gamma_{2,4}}{\gamma_{0}\partial \rho}$ & 1.05 & 1.06\\

$\sfrac{\partial \gamma_{3,5}}{\gamma_{0}\partial \rho}$ & 0.32 & 0.29\\

$\sfrac{\partial \gamma_{1,4}}{\gamma_{0}\partial \rho}$ & -0.84 & -0.82\\
\hline\hline
\multirow{2}{*}{Sensitivity coefficients} & Simulated  & Experimental \\
 & \textcolor{black}{$\times 10^{6}$~µm$^{2}$} & \textcolor{black}{$\times 10^{6}$~µm$^{2}$} \\
\hline\hline
$\sfrac{\partial^{2} \overline{\gamma}}{\gamma_{0}\partial \rho^{2}}$ & $0.18\textcolor{black}{\times}10^{-3}$& $3.24\textcolor{black}{\times}10^{-3}$\\
\hline
\end{tabular}
\end{table}
Although, there is a good agreement for $\frac{1}{\gamma_{0}}\frac{\partial \gamma_{i,j}}{\partial \rho}$, there exists a significant discrepancy of factor of $18$ in $\frac{1}{\gamma_{0}}\frac{\partial^{2} \overline{\gamma}}{\partial \rho^{2}}$. Moreover, the accepted tolerance for the displacement $\rho$ is found to be
\begin{eqnarray}
u^{2}(\rho) & < &  \frac{10^{-8}}{|\frac{1}{\gamma_{0}}\frac{\partial^{2} \overline{\gamma}}{\partial \rho^{2}} |} \Rightarrow  u(\rho)<7.4~\text{µm} \label{eq:rhotol}
\end{eqnarray}
when the simulated coefficient is used. This tolerance is not at the sub-micron level as expected and is not the prevailing error in the standard, as it was thought to be initially. Consequently, we investigated further this matter to provide a more comprehensive understanding of the experimental results. 

\subsection{Rotation of the movable electrode}
\begin{figure}[t]
  \center
  \includegraphics[width=3.4in]{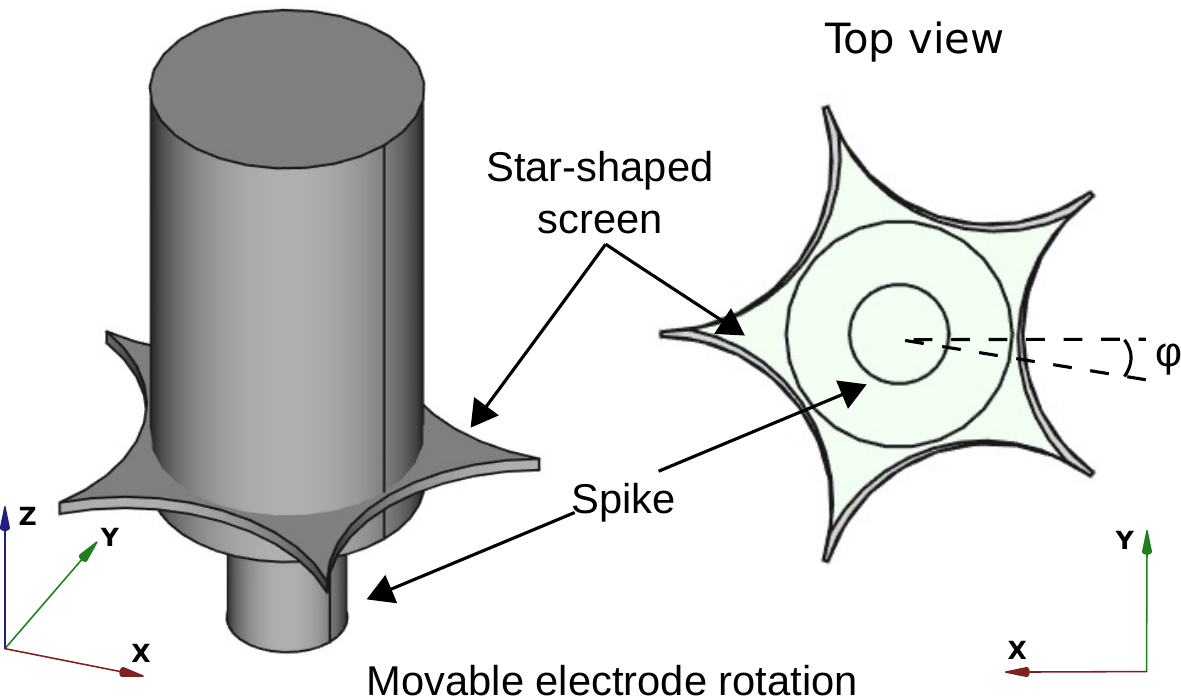}
  \caption{(left) The movable electrode with a star-shaped screen and a spike and (right) its axial rotation by an angle $\varphi$. This rotation occurs when the movable electrode is moved from position $L_{1}$ to $L_{2}$.}
  \label{fig:etoileFigure}
\end{figure}
\begin{figure}[t]
  \center
  \includegraphics[width=3.4in]{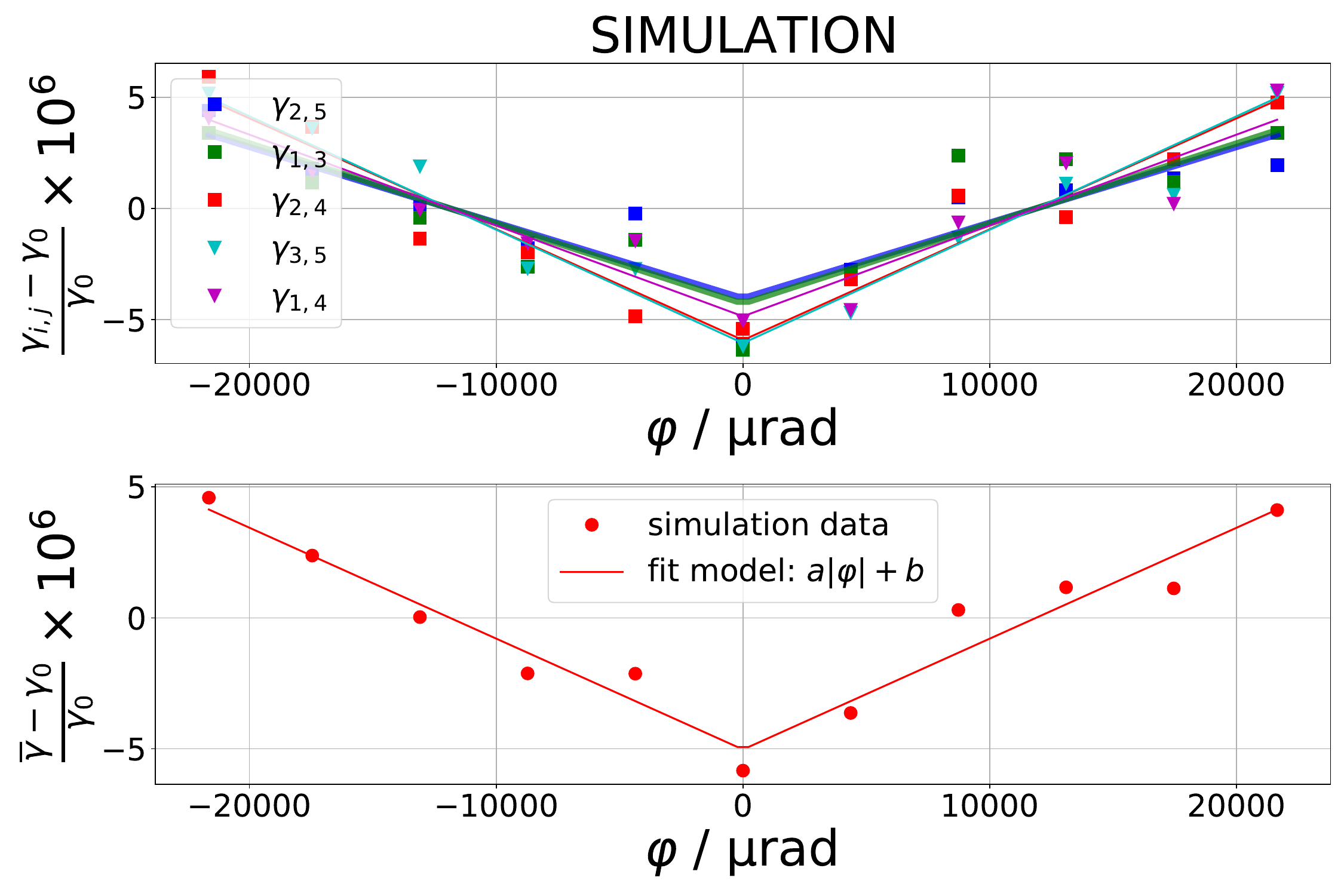}
  \caption{(Top plot) Relative changes in $\gamma_{i,j}$ as a function of the movable electrode rotation $\varphi$, and (Bottom plot) relative changes in their average value. For this simulation's outcomes, the mobile electrode features a star-shaped screen, depicted in Fig.~\ref{fig:etoileFigure}.}
  \label{fig:spikerotFigure}
\end{figure}
So far, simulations have utilized a basic form of the movable guard: a cylinder with a spike at the tip, depicted in Fig.~\ref{fig:systemFigure}. However, in practice, the movable electrode of the TLCC also includes a star-shaped screen, as shown in Fig.~\ref{fig:etoileFigure}. In experimental conditions, the displacement of the movable guard by $\rho$ may cause it to rotate \textcolor{black} {about z-axis, resulting in a radial tilt that cannot be directly measured. Tilts in tangential directions (rotations about x and y-axis) are negligibly small, as seen by the optical interferometer.} Consequently, the rotation of the star-shaped screen by an angle $\varphi$ \textcolor{black}{about z-axis} induces a change in $\overline{\gamma}$. This scenario is simulated and its result is depicted in Fig.~\ref{fig:spikerotFigure}, along with the corresponding sensitivity coefficients outlined in Table~\ref{tab:table3}. A regression model of the form $a|\varphi|+b$ was used to determine these coefficients. Notably, the observed sensitivity coefficients, $\frac{1}{\gamma_{0}}\frac{\partial \gamma_{i,j}}{\partial |\varphi|}$ and $\frac{1}{\gamma_{0}}\frac{\partial \overline{\gamma}}{\partial |\varphi|}$, are of the same order, which is in strong contrast to previous cases where there was a difference of one or two orders of magnitude between them. This suggests that the observed variation in $\overline{\gamma}$ in the experimental data from the previous section is primarily due to the rotation of the movable electrode, whereas the variations in $\gamma_{i,j}$ are linked to the displacement of the same electrode.
\begin{table}[h]
\caption{Sensitivity coefficients of the trajectory error ($\varphi$) \label{tab:table3}}
\renewcommand\arraystretch{1.3}
\centering
\begin{tabular}{|c||c|}
\hline
\multirow{2}{*}{Sensitivity coefficients} & Simulated \\
 & \textcolor{black}{$\times 10^{6}$~µrad} \\
\hline\hline
$\sfrac{\partial \gamma_{2,5}}{\gamma_{0}\partial |\varphi|}$ &  $3.4\textcolor{black}{\times}10^{-4}$\\

$\sfrac{\partial \gamma_{1,3}}{\gamma_{0}\partial |\varphi|}$ & $3.6\textcolor{black}{\times}10^{-4}$ \\

$\sfrac{\partial \gamma_{2,4}}{\gamma_{0}\partial |\varphi|}$ & $4.9\textcolor{black}{\times}10^{-4}$ \\

$\sfrac{\partial \gamma_{3,5}}{\gamma_{0}\partial |\varphi|}$ & $5.1\textcolor{black}{\times}10^{-4}$ \\

$\sfrac{\partial \gamma_{1,4}}{\gamma_{0}\partial |\varphi|}$ & $4.1\textcolor{black}{\times}10^{-4}$ \\
\hline\hline
$\sfrac{\partial \overline{\gamma}}{\gamma_{0}\partial |\varphi|} $ & $4.2\textcolor{black}{\times}10^{-4}$ \\
\hline
\end{tabular}
\end{table}

The acceptable tolerance for the rotation angle $\varphi$ of the movable electrode can be assessed as follows:
\begin{eqnarray}
u(\varphi) & < &  \frac{10^{-8}}{|\frac{1}{\gamma_{0}}\frac{\partial \overline{\gamma}}{\partial \varphi} |} \approx 24~\text{µrad} \label{eq:tolvarphi}
\end{eqnarray}

\subsection{Simultaneous rotation and displacement of the movable electrode} \label{ssec:num3}
The experimental data in Fig.~\ref{fig:trajerrexpFigure} can now be more thoroughly explored in terms of the simultaneous rotation and displacement of the movable electrode. In the experimental setting, the rotation angle $\varphi$ is not directly measurable. We will assume a one-to-one relationship between the rotation angle $\varphi$ and the displacement $\rho$, meaning their ratio $\lambda=\frac{\varphi}{\rho}$ remains constant. 
\begin{figure}[t]
  \center
  \includegraphics[width=3.4in]{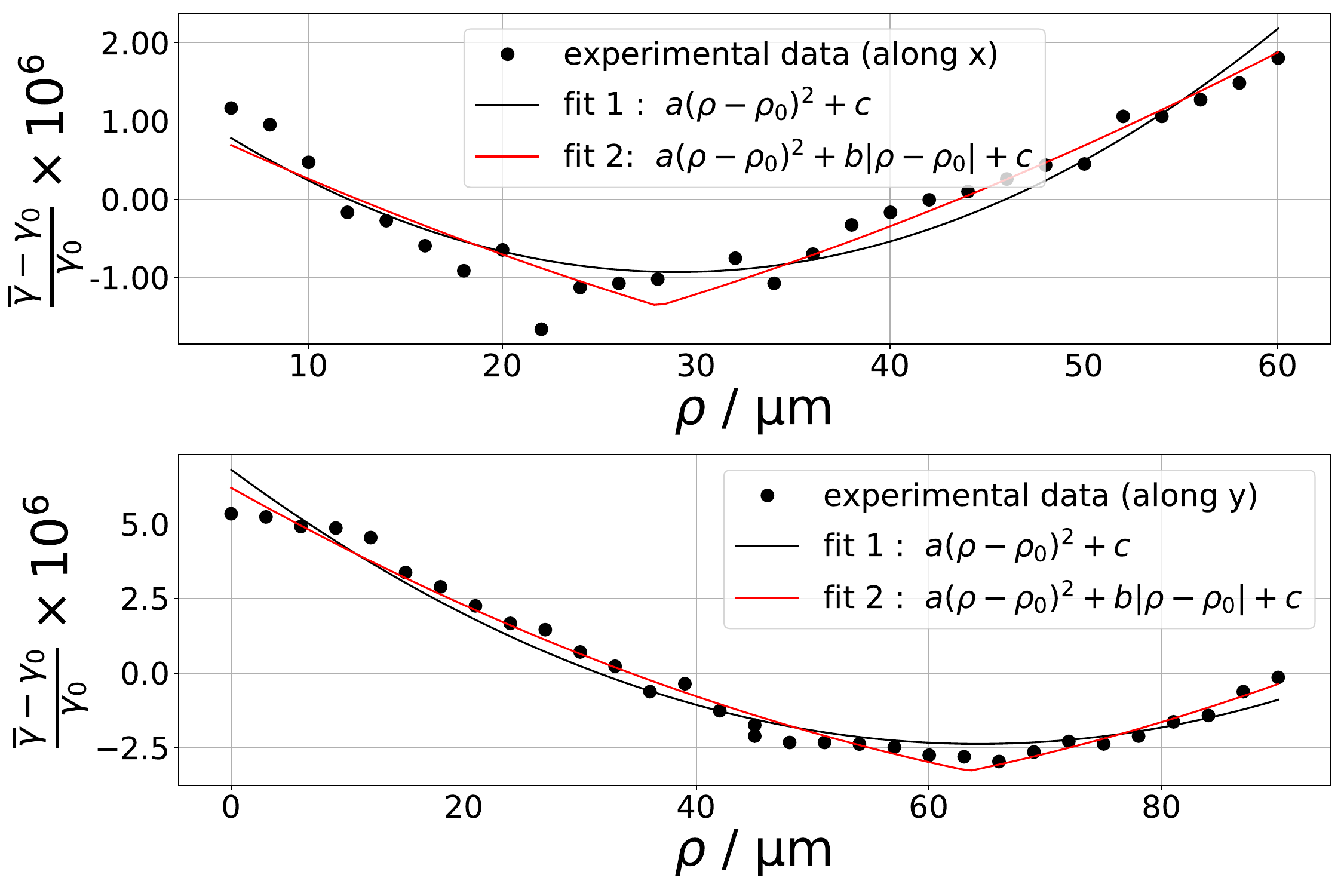}
  \caption{The experimental data of relative changes in $\overline{\gamma}$ as a function of the movable electrode displacement $\rho$ (top plot) along x-axis and (bottom plot) along y-axis, and the fit of each dataset with a parabola representing the sole displacement of the movable electrode (black curve) and fit with a combined model representing both its rotation and displacement (red curve)}
  \label{fig:expfitmodelFigure}
\end{figure}

A regression model of the form $a(\rho-\rho_{0})^{2}+\frac{b}{\lambda}|\varphi-\varphi_{0}|+c$, incorporating both the rotation $\frac{b}{\lambda}|\varphi-\varphi_{0}|=b|\rho-\rho_{0}|$ and the displacement $a(\rho-\rho_{0})^{2}$ effects of the movable electrode, is fitted to the experimental data of $\frac{\overline{\gamma}-\gamma_{0}}{\gamma_{0}}$. The determined coefficients provide insight into the relative importance of each contribution. The results are depicted in Fig.~\ref{fig:expfitmodelFigure}, where the top plot represents the electrode displacement along the x-axis (previously presented in Fig.~\ref{fig:trajerrexpFigure}), and the bottom plot shows the displacement along the y-axis, orthogonal to the x-axis in the horizontal plane. The quadratic fit is also included in these plots for comparison with the combined model fit. 
\begin{table}[h]
\caption{Identification of the trajectory error in the LNE TLCC \label{tab:table4}}
\renewcommand\arraystretch{1.3}
\centering
\begin{tabular}{|c||c|c|}
\hline
\multirow{3}{*}{Experiments} & Rotation Sensitivity  & Displacement Sensitivity \\
 & $\frac{1}{\gamma_{0}}\frac{\partial \overline{\gamma}}{\partial |\rho|} $ & $\frac{1}{\gamma_{0}}\frac{\partial^{2} \overline{\gamma}}{\partial \rho^{2}} $\\
 & \textcolor{black}{$\times 10^{6}$~µm} & \textcolor{black}{$\times 10^{6}$~µm$^{2}$}\\
\hline\hline
$\rho$ along x-axis &  $7.5\textcolor{black}{\times}10^{-2}$& $8.1\textcolor{black}{\times}10^{-4}$\\
$\rho$ along y-axis & $8.1\textcolor{black}{\times}10^{-2}$& $1.1\textcolor{black}{\times}10^{-3}$\\
\hline
\end{tabular}
\end{table}
The determined coefficients, outlined in Table~\ref{tab:table4}, reveal that the primary contribution to the trajectory error arises from the rotation of the movable electrode. Furthermore, the displacement sensitivity coefficient $\frac{1}{\gamma_{0}}\frac{\partial^{2} \overline{\gamma}}{\partial \rho^{2}}$ closely approximates its counterpart determined through simulations in the previous section in Table~\ref{tab:table2}. Consequently, the accepted tolerance for the displacement $\rho$ can be reevaluated as:
\begin{eqnarray}
u(\rho) & < &  \frac{10^{-8}}{|\frac{1}{\gamma_{0}}\frac{\partial \overline{\gamma}}{\partial |\rho|} |} \Rightarrow  u(\rho)<120\text{ nm} \label{eq:rhotolrev}
\end{eqnarray}
This revised tolerance is much more stringent than the result in equation (\ref{eq:rhotol}). Using this revised tolerance $u(\rho)$ and the previously established tolerance $u(\varphi)$ in (\ref{eq:tolvarphi}), the relation between $\varphi$ and $\rho$ can be estimated as: $\lambda=\frac{\varphi}{\rho}=\frac{24}{0.12} =200\ \frac{\text{µrad}}{\text{µm}}$. 

\section{Resulting Uncertainties in the new TLCC}
In this section, we will briefly discuss the actual uncertainty contributions from the parallelism and trajectory errors in the second generation TLCC at LNE and compare them with the acceptable tolerances established via the FEM simulations so far. Regarding the trajectory error, as detailed in \cite{thevenot2023progress}, the displacement of the movable guard is controlled at the level of $80$ nm, resulting in an uncertainty $u(\rho)$ of similar magnitude. This uncertainty is compared with the acceptable tolerance $u(\rho)<120$ nm in equation (\ref{eq:rhotolrev}). On the other hand, concerning the parallelism error, the uncertainty of the tilt angle $u(\alpha)$ of the five electrodes were determined from the linear regression and its uncertainty to the experimental data in Fig. 9 of \cite{thevenot2023progress}. The resulting uncertainty of the tilt angle of every static electrode is about $u(\alpha)=0.25$~µrad to be compared with the acceptable tolerance $u(\alpha)<0.46$~µrad in equation (\ref{eq:alphatol}). 

\textcolor{black}{Using the experimental estimates for $u(\alpha)$ and $u(\rho)$ as given above, along with their respective sensitivity coefficients obtained from the present analysis, the uncertainty contributions from these mechanical imperfections to $\frac{u(\overline{\gamma})}{\gamma_{0}}$  can be estimated using equation (\ref{eqsenscoeff1}). The resulting combined relative uncertainty is about $0.86\times10^{-8}$, representing at least a factor of 4 improvement compared to the first generation TLCC \cite{trapon2003determination}. Additionally, the uncertainty contribution from the bridge ratio $\frac{u(r)}{r}$ has been improved to $0.5\times10^{-8}$ \cite{thevenot2023progress}. These accomplishments mark a highly promising achievement with the new generation calculable capacitor meeting or surpassing the target uncertainty of one part in $10^{8}$ for $\frac{u(C)}{C}$.}

\section{Conclusion}

Imperfections in the TLCC are limiting factors in the realization of the farad with a target uncertainty of one part in $10^{8}$. Specifically, the presence of parallelism defects in static electrodes and errors in the trajectory of the movable electrode emerges as main limitations in the LNE TLCC standard. Assessing their impact on the TLCC uncertainty serves a dual purpose: firstly, establishing their contribution within the overall uncertainty budget, and secondly, identifying targeted improvements to mitigate these defects.

This study demonstrates that employing FEM simulations proves crucial in understanding the real-world effects of these imperfections. The results obtained for the mentioned prominent errors through FEM analysis align with experimental outcomes, affirming its capability to simulate and predict the TLCC behavior independently. Beyond mere simulation, FEM also helps in interpreting complex experimental data.

In summary, this research emphasizes the utility of FEM simulations as a valuable tool for evaluating the effects of the TLCC imperfections, especially when these imperfections are challenging to access experimentally.


\section*{Acknowledgment}
We thank Dr. François Piquemal, Dr. Wilfrid Poirier and Dr. Sophie Djordjevic for comments that greatly improved the manuscript.


\bibliographystyle{IEEEtran.bst}
%





\newpage

{\appendix[Relation between $\overline{e}$ and $\frac{u(\overline{\gamma})}{\gamma_{0}}$]
When there is a relatively small imperfection of the system, the five capacitances per unit length are given by:
\begin{eqnarray*}
\gamma_{i,j} & = &  \gamma_{0} + \Delta\gamma_{i,j}
\end{eqnarray*}
 such that $\Delta\gamma_{i,j}<<\gamma_{0}$. Their average is then:
\begin{eqnarray*}
\overline{\gamma} & = &  \gamma_{0} + \frac{1}{5}\sum\Delta\gamma_{i,j}\\
                  & = &  \gamma_{0} + \Delta\overline{\gamma}
\end{eqnarray*}
Two categories of imperfections need to be addressed: those where $\Delta\overline{\gamma}=0$ and those where $\Delta\overline{\gamma}\neq 0$. The former type, not pertinent to our current discussion, does not introduce significant uncertainties in $C$ at the first order. These imperfections typically stem from the symmetrical characteristics of the static electrodes' cross-section, including angular and radial positioning or differences in diameter. For our purposes, we will focus on the latter type of imperfection, which pertains to errors in coaxiality or cylindricity of the volume enclosed by the five static electrodes.

Then, the errors in the Lampard's equations can be written as:
\begin{eqnarray*}
e_{1} & = & e^{-\frac{\pi}{\epsilon_{0}}2\gamma_{0}}e^{-\frac{\pi}{\epsilon_{0}}(\Delta\gamma_{1,3}+\Delta\gamma_{1,4})}+ e^{-\frac{\pi}{\epsilon_{0}}\gamma_{0}}e^{-\frac{\pi}{\epsilon_{0}}\Delta\gamma_{2,5}} - 1 \\
e_{2} & = & e^{-\frac{\pi}{\epsilon_{0}}2\gamma_{0}}e^{-\frac{\pi}{\epsilon_{0}}(\Delta\gamma_{2,4}+\Delta\gamma_{2,5})}+ e^{-\frac{\pi}{\epsilon_{0}}\gamma_{0}}e^{-\frac{\pi}{\epsilon_{0}}\Delta\gamma_{1,3}} - 1 \\
e_{3} & = & e^{-\frac{\pi}{\epsilon_{0}}2\gamma_{0}}e^{-\frac{\pi}{\epsilon_{0}}(\Delta\gamma_{3,5}+\Delta\gamma_{1,3})}+ e^{-\frac{\pi}{\epsilon_{0}}\gamma_{0}}e^{-\frac{\pi}{\epsilon_{0}}\Delta\gamma_{2,4}} -  1 \\
e_{4} & = & e^{-\frac{\pi}{\epsilon_{0}}2\gamma_{0}}e^{-\frac{\pi}{\epsilon_{0}}(\Delta\gamma_{1,4}+\Delta\gamma_{2,4})}+ e^{-\frac{\pi}{\epsilon_{0}}\gamma_{0}}e^{-\frac{\pi}{\epsilon_{0}}\Delta\gamma_{3,5}} - 1 \\
e_{5} & = & e^{-\frac{\pi}{\epsilon_{0}}2\gamma_{0}}e^{-\frac{\pi}{\epsilon_{0}}(\Delta\gamma_{2,5}+\Delta\gamma_{3,5})}+e^{-\frac{\pi}{\epsilon_{0}}\gamma_{0}} e^{-\frac{\pi}{\epsilon_{0}}\Delta\gamma_{1,4}} -  1
\end{eqnarray*}
and carrying out Taylor expansion up to the first order for $\frac{\pi}{\epsilon_{0}}\Delta\gamma_{i,j}$: 
\footnotesize
\begin{eqnarray*}
e_{1} & = & e^{-\frac{\pi}{\epsilon_{0}}2\gamma_{0}}(1-\frac{\pi}{\epsilon_{0}}(\Delta\gamma_{1,3}+\Delta\gamma_{1,4}))+ e^{-\frac{\pi}{\epsilon_{0}}\gamma_{0}}(1-\frac{\pi}{\epsilon_{0}}\Delta\gamma_{2,5}) - 1 \\
e_{2} & = & e^{-\frac{\pi}{\epsilon_{0}}2\gamma_{0}}(1-\frac{\pi}{\epsilon_{0}}(\Delta\gamma_{2,4}+\Delta\gamma_{2,5}))+ e^{-\frac{\pi}{\epsilon_{0}}\gamma_{0}}(1-\frac{\pi}{\epsilon_{0}}\Delta\gamma_{1,3}) - 1 \\
e_{3} & = & e^{-\frac{\pi}{\epsilon_{0}}2\gamma_{0}}(1-\frac{\pi}{\epsilon_{0}}(\Delta\gamma_{3,5}+\Delta\gamma_{1,3}))+ e^{-\frac{\pi}{\epsilon_{0}}\gamma_{0}}(1-\frac{\pi}{\epsilon_{0}}\Delta\gamma_{2,4}) -  1 \\
e_{4} & = & e^{-\frac{\pi}{\epsilon_{0}}2\gamma_{0}}(1-\frac{\pi}{\epsilon_{0}}(\Delta\gamma_{1,4}+\Delta\gamma_{2,4}))+ e^{-\frac{\pi}{\epsilon_{0}}\gamma_{0}}(1-\frac{\pi}{\epsilon_{0}}\Delta\gamma_{3,5}) - 1 \\
e_{5} & = & e^{-\frac{\pi}{\epsilon_{0}}2\gamma_{0}}(1-\frac{\pi}{\epsilon_{0}}(\Delta\gamma_{2,5}+\Delta\gamma_{3,5}))+e^{-\frac{\pi}{\epsilon_{0}}\gamma_{0}}(1-\frac{\pi}{\epsilon_{0}}\Delta\gamma_{1,4}) -  1
\end{eqnarray*}
\normalsize
Using the property $e^{-\frac{\pi}{\epsilon_{0}}2\gamma_{0}}+e^{-\frac{\pi}{\epsilon_{0}}\gamma_{0}}-1=0$ and averaging the five equations we find:
\begin{eqnarray*}
\overline{e} & = & -2\frac{\pi}{\epsilon_{0}}\Delta \overline{\gamma} e^{-\frac{\pi}{\epsilon_{0}}2\gamma_{0}}-\frac{\pi}{\epsilon_{0}}\Delta \overline{\gamma} e^{-\frac{\pi}{\epsilon_{0}}\gamma_{0}} \\
            & = & -\frac{\pi}{\epsilon_{0}}\Delta \overline{\gamma} (1+e^{-\frac{\pi}{\epsilon_{0}}2\gamma_{0}}) \\
            & = & -\frac{\pi}{\epsilon_{0}}\gamma_{0}(1+e^{-2\frac{\pi}{\epsilon_{0}}\gamma_{0}})\frac{\Delta \overline{\gamma}}{\gamma_{0}} \\
            & = & -\ln \frac{2}{\sqrt{5}-1} (1+e^{-2\ln \frac{2}{\sqrt{5}-1} })\frac{\Delta \overline{\gamma}}{\gamma_{0}} \\
            & = & -\frac{1}{\kappa}\frac{\Delta \overline{\gamma}}{\gamma_{0}}
\end{eqnarray*}
Since $\Delta \overline{\gamma}$ cannot be experimentally determined accurately, it is instead considered as part of the uncertainty in $\overline{\gamma}$ and
consequently:
\begin{eqnarray*}
\overline{e} & = &  -\frac{1}{\kappa}\frac{u( \overline{\gamma})}{\gamma_{0}}
\end{eqnarray*}

}

\end{document}